\documentclass[useAMS,usenatbib]{mn2e}

\usepackage{graphicx}

\usepackage{booktabs}

\title[
Multiple Shells in IC\,418]{
Discovery of Multiple Shells Around the Planetary Nebula IC\,418}

\author[G. Ramos-Larios et al.] 
{G. Ramos-Larios$^{1}$\thanks{E-mail:
gerardo@astro.iam.udg.mx (GRL)}
R. V\'azquez$^{2}$, M.A. Guerrero$^{3}$, L. Olgu\'{\i}n$^{4}$, R.A. Marquez-Lugo$^{1}$ \newauthor and H. Bravo-Alfaro$^{5}$\\
$^{1}$Instituto de Astronom\'{\i}a y Meteorolog\'{\i}a, 
Av.\ Vallarta No.\ 2602, Col.\ Arcos Vallarta, C.P. 44130 Guadalajara, Jalisco, Mexico.\\
$^{2}$Instituto de Astronom\'{\i}a, Universidad Nacional Aut\'onoma de M\'exico, Apdo. Postal 877, 22800 Ensenada, B.C., Mexico\\
$^{3}$Instituto de Astrof\'{\i}sica de Andaluc\'{\i}a, IAA-CSIC, C/ Glorieta de la Astronom\'{\i}a s/n, 18008 Granada, Spain\\
$^{4}$Dpto. de Investigaci\'on en F\'{\i}sica, Universidad de Sonora, Blvd. Rosales-Colosio, Ed. 3H, 83190, Hermosillo, Sonora, Mexico\\
$^{5}$Departamento de Astronom\'{\i}a, Universidad de Guanajuato, Apdo. Postal 144, Guanajuato 36000, Mexico}

\begin{document}

\date{Received 2011 September 27; in original form 2011 June 16}

\pagerange{\pageref{firstpage}--\pageref{lastpage}} \pubyear{2012}

\maketitle

\label{firstpage}

\begin{abstract}

We have analysed optical, near-, and mid-IR images of the bright planetary 
nebula (PN) IC\,418.  
These images probe unambiguously for the first time a number of 
low surface brightness structures and shells around the bright main nebula, including radial filaments or rays, a system of three concentric 
rings, and two detached haloes with sizes $\sim$150\arcsec\ and 
220\arcsec$\times$250\arcsec, respectively.  
The main nebula is slightly off-centered with respect to the elliptical outer halo. The time-lapse between the two haloes is 10,000--50,000\,yr, whereas 
the time-lapse between the three concentric rings is $\sim$630\,yr. 
We emphasize the advantages of near- and mid-IR imaging for the detection 
of faint structures and shells around optically bright nebulae.

\end{abstract}


\section{Introduction}

IC\,418 is a bright elliptical planetary nebula (PN) whose physical 
structure and emission properties have been subject of numerous 
observational studies and theoretical modeling \citep[e.g.,][]{
1990AA...234..454P,1992ApJ...385..255Z,1994PASP..106..745H,
1996AA...313..234M,2009AA...507.1517M}.  
The nebula is described as a high density ellipsoidal shell with 
chemical abundances typical of type II PNe. 
The distance to IC\,418 is 1.3$\pm$0.4 kpc as determined from VLA
parallax expansion observations spanning more than 20 years in time
\citep{2009AJ....138...46G}, making of IC\,418 one of the few PNe with a reliable distance measurement.
Its central star has a relatively low effective temperature of 39,000~K 
\citep{2004AA...423..593P}, and a surface gravity $\log(g)$ of 3.70 
\citep{2004AA...419.1111P}, implying an early evolutionary stage.

Whereas the overall morphology of IC\,418 seems simple, the detailed view 
of this bright nebula offered by \emph{HST} WFPC2 images \citep{ST98} 
reveals an intricate cyclic pattern that has dubbed it as the ``Spirograph 
Nebula''.  
The origin of this complex pattern is intriguing: magnetic fields 
have been claimed to be responsible for these structures \citep{HM05}, 
although the action of its variable stellar wind, revealed by 
changes in the P~Cygni profiles of high excitation lines detected 
by \emph{FUSE} \citep[][Guerrero \& De Marco, in preparation]
{2012arXiv1202.6544P} and the 
photometric variability of the central star \citep{Handler_etal97}, 
can also be linked to the formation of this singular pattern.

The Zanstra temperatures of IC\,418 have been reported to be 
$T_{\rm Z}$(H~{\sc i}) = 38,000~K and $T_{\rm Z}$(He~{\sc ii}) = 44,000~K
\citep[e.g.,][]{2003MNRAS.344..501P}, which are in agreement with the
temperature estimate obtained from the energy balance method
\citep{2004AA...423..593P}.  
A small difference between the H~{\sc i} and He~{\sc ii} Zanstra
temperatures is traditionally interpreted as proof of the large
optical thickness of the nebula to H-ionising radiation.  
Therefore, significant amounts of atomic and molecular material, as
well as dust, may survive outside the ionisation bound optical nebula.

Detached low surface brightness shells, i.e., haloes, are typically found 
around PNe \citep[e.g.,][]{1987ApJS...64..529C}.  
These haloes have been associated to the final mass-loss episodes that 
occur in the late phases of the AGB \citep{1992ApJ...392..582B}, and 
therefore they can provide information on the late AGB evolution 
\citep[e.g.,][]{FVB94,1995ApJ...452..286S,Hajian_etal97}.  
The identification and detection of low surface brightness shells around 
bright PNe is hampered by light from the inner shell dispersed within the 
optical system used for the observations \citep[see][for a thorough 
description of the observational problems associated to such studies]
{2003MNRAS.340..417C}. 

An early investigation of IC\,418 by \citet{1986ApSS.122...81P} proposed that the 
unusually extended near-IR emission may arise from high temperature,
small-sized dust grains.  It was also proposed that scattering could
result in a $\sim$40\arcsec\ in diameter optical halo \citep[e.g.,
][]{1990ApSS.171..173P,1990AA...234..454P} The presence of this halo
and the high temperature of the dust have been later confirmed by the
analysis of \emph{J$-$H} and \emph{H$-$K$_s$} maps by
\citet{2005MNRAS.364..849P} who also suggested that the grains are
probably mixed with a dilute ionised gas.  Recently,
\citet{2011IAUS283} have modelled the IR dust emission from 2 to 200
$\mu$m and concluded that the dust is carbon rich.

Meanwhile \citet{1990MNRAS.242..457M} have argued that such outer shell is not real, but it should be attributed to a mix of Galactic background emission and extended King seeing function. On the other hand, \citet{1989ApJ...340..932T} detected a region of H~{\sc i} emission whose spatial extent seems to be coincident with the region described by \citet{1990AA...234..454P}, whereas  \citet{1995MNRAS.273..790G} found no evidence for molecular emission.
These results suggest that the material outside the bright inner shell of IC\,418 is photo-dissociated.  
Indeed, \citet{1996AA...313..234M}, using mid-IR and radio images, suggested that the inner shell of IC\,418 is surrounded by a low density ionised region with radius $\sim$20\arcsec\ that is enclosed into an atomic neutral halo with an outer radius of 90\arcsec. 
Additional evidence for the occurrence of a halo of ionised material is provided by the near-IR spectroscopy of this region reported by \citet{1999ApJS..124..195H} who detected emission in the Pa$\beta$ $\lambda$1.2817 
$\mu$m, Br8 H~{\sc i} $\lambda$1.9446 $\mu$m, He~{\sc i} $\lambda$2.058 $\mu$m, and Br$\gamma$ $\lambda$2.1658 $\mu$m lines.  
Unfortunately, the exact location of the slit (east of the main nebula) and extent of the emission are not described, hence the nature of the outer shell mentioned by these authors and its spatial extent remains uncertain.

In order to assess the presence of material outside the bright ionised shell of IC\,418 and to investigate its nature, we have acquired new broad-band deep near-IR $JHK$ images, and long slit intermediate- and high-dispersion spectroscopic observations that have been examined in conjunction with archival mid-IR \emph{WISE} images.  
The observations and archival data are presented in \S\ref{sec_obs} and the results are described in \S\ref{sec_res}. 
The discussion and final summary are presented in 
\S\ref{sec_dis} and \S\ref{sec_con}, respectively.

\begin{table}\centering
\setlength{\columnwidth}{0.2\columnwidth}
\setlength{\tabcolsep}{1.30\tabcolsep}
\caption{NIR Imaging of IC\,418}
\begin{tabular}{llrrr}
\hline
\multicolumn{1}{l}{Telescope} & 
\multicolumn{1}{c}{Filter} & 
\multicolumn{1}{c}{$\lambda_{c}$} & 
\multicolumn{1}{c}{$\Delta\lambda$} & 
\multicolumn{1}{c}{Exp.\ Time} \\
\multicolumn{1}{c}{} & 
\multicolumn{1}{c}{} & 
\multicolumn{1}{c}{($\mu$m)} & 
\multicolumn{1}{c}{($\mu$m)} & 
\multicolumn{1}{c}{(s)} \\ 
\hline

2.1m OAN-SPM &  J  &  1.275 &  0.282 & 600~~~~ \\
2.1m OAN-SPM &  H  &  1.672 &  0.274 & 400~~~~ \\
2.1m OAN-SPM &  K' &  2.124 &  0.337 & 300~~~~ \\
\hline
3.5m TNG-ORM &  J  &  1.270 &  0.300 &  50~~~~ \\
    \hline
  \end{tabular}
\vspace{0.4cm}
\end{table}

\section[]{Observations and Archival Data}\label{sec_obs}

\subsection{Optical imaging}

Narrow-band \emph{HST} WFPC2 images of IC\,418 in the H$\alpha$, 
[N~{\sc ii}] $\lambda$6583, and [O~{\sc iii}] $\lambda$5007 emission 
lines were retrieved from HLA, the Hubble Legacy Archive\footnote{
http://hla.stsci.edu/} 
at the Space Telescope Science Institute\footnote{
STScI is operated by the Association of Universities for Research in 
Astronomy, Inc., under NASA contract NAS5-26555.} 
(Prop.\ ID: 6353, PI: R. Sahai, and Prop.\ ID: 8773, PI: A. Hajian).  
  
The 888~s H$\alpha$ image was obtained from eight individual 
exposures acquired through the F656N filter (pivot wavelength 
$\lambda_{\rm p}$=6563.8\,\AA and band-width $\Delta\lambda$=21.5\,\AA), whereas 
the 700~s [N~{\sc ii}] image and the 600~s [O~{\sc iii}] were 
obtained from three individual exposures acquired using the F658N 
($\lambda_{\rm p}$=6590.8\,\AA, $\Delta\lambda$=28.5\,\AA) and F502N 
($\lambda_{\rm p}$=5012.4\,\AA, $\Delta\lambda$=26.9\,\AA) filters, 
respectively.

\subsection{Near-IR imaging}

Broad-band near-IR images of IC\,418 were acquired on October 23, 2005 
using the CAMILA infrared camera \citep{1994RMxAA..29..197C} mounted 
with f/4.5 focal reducing optics at the f/13.5 primary focus of the 2.1m 
telescope at the Observatorio Astron\'omico Nacional, San Pedro 
M\'artir (OAN-SPM, Baja California, Mexico). 
CAMILA has a 256$\times$256 pixel NICMOS3 detector with a plate scale of 
0\farcs85~pixel$^{-1}$ and a field of view (FoV) of 3\farcm6$\times$3\farcm6. 
Further details of the observations are provided in Table~1.  

The source was observed following a typical on-off sequence and the 
resulting individual frames were reduced following standard procedures. The seeing was $\simeq$2\farcs6.

A broad-band $J$ image of the central source of IC\,418 was also obtained 
on October 13, 2008 using NICS, the Near Infrared Camera Spectrometer 
\citep{2001AA...378..722B}, at the Cassegrain focus of the 3.5-m Telescopio 
Nazionale Galileo (TNG) on Roque de Los Muchachos Observatory (ORM, La 
Palma, Spain).  
NICS is a multimode instrument for near-IR observations in the range 
0.9--2.5 $\mu$m that employs a Rockwell 1024$\times$1024 HgCdTe Hawaii 
array.  
The large field (LF) mode was used, providing a plate scale of 
0\farcs25~pixel$^{-1}$ and a FoV of 4\farcm2$\times$4\farcm2. 
Five 10 s exposures were secured for a total effective 
exposure time of 50 s (Table~1).   

The NICS data reduction was carried out using {\small SNAP} ~(Speedy Near-IR data 
Automatic reduction Pipeline), a pipeline for the automatic reduction of 
near-IR data that uses pieces of existing software such as {\small IRDR}\footnote{
IRDR, the Infra-Red Data Reduction, is a C library and set of 
stand-alone C programs and perl scripts for processing IR 
imaging data that is distributed by the Institute of Astronomy 
(IoA), at the University of Cambridge.}, 
{\small IRAF}\footnote{
{\small IRAF}, the Image Reduction and Analysis Facility, is distributed 
by the National Optical Astronomy Observatory, which is operated 
by the Association of Universities for Research in Astronomy 
(AURA) under cooperative agreement with the National Science 
Foundation.}, 
{\small Sextractor}, and {\small Drizzle}\footnote{
{\small Drizzle} is available as an {\small IRAF} task as part of the Space Telescope 
Science Data Analysis System ({\small STSDAS}) package and can be retrieved 
from the Space Telescope Science Institute (STScI) web site.}. 
The reduction performed by {\small SNAP} includes additional non-standard steps such 
as cross-talk correction, double-pass sky subtraction, and field distortion 
correction.    
The spatial resolution, as determined from the FWHM of stars in the FoV, 
is $\simeq$0$\farcs$7.

\subsection{Mid-IR imaging}

Wide-field Infrared Survey Explorer, \emph{WISE} \citep{2010AJ....140.1868W}, 
images of IC\,418 were retrieved from the NASA/IPAC Infrared Science 
Archive (IRSA). 
\emph{WISE} is a NASA Explorer mission that surveys the entire sky 
at 3.4, 4.6, 12, and 22 $\mu$m, the so-called W1 through W4 bands, with 5$\sigma$ point source sensitivities better than 0.08, 0.11, 1, and 6 mJy, respectively.

The 40-cm telescope uses HgCdTe and Si:As detectors arrays with a 
plate scale of 2\farcs75~pixel$^{-1}$.    
The data were downloaded from the \emph{WISE} first data release 
that includes a preliminary catalogue of sources and an image 
atlas based on early data that covers more than 55\% of the sky.  
The angular resolution in the four bands is 6\farcs1, 6\farcs4, 
6\farcs5, and 12\farcs0, respectively, and the astrometric accuracy 
for bright sources is better than 0\farcs15.

\begin{figure}
\includegraphics[width=0.9\linewidth]{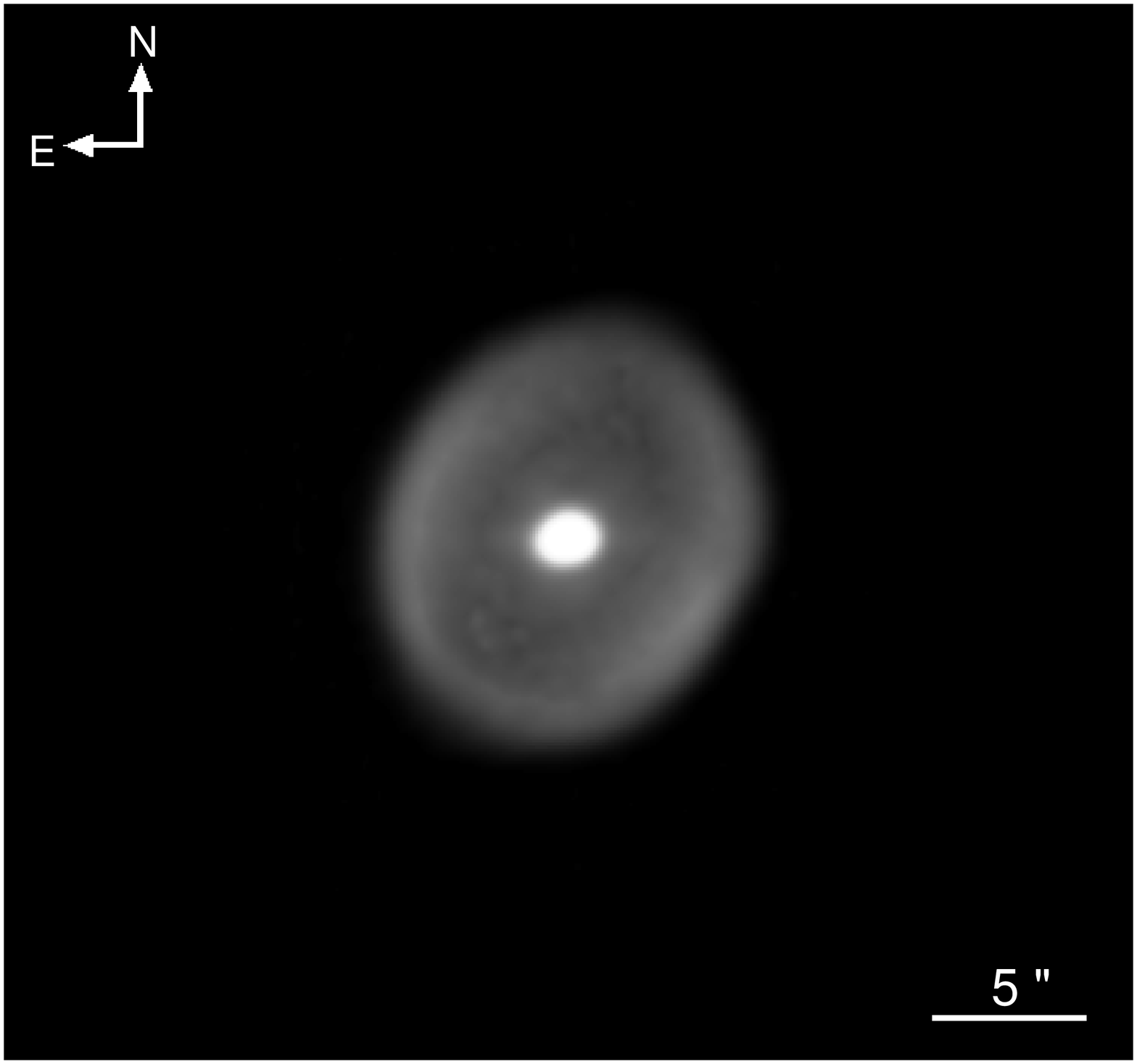}
\vskip .1in
\includegraphics[width=0.9\linewidth]{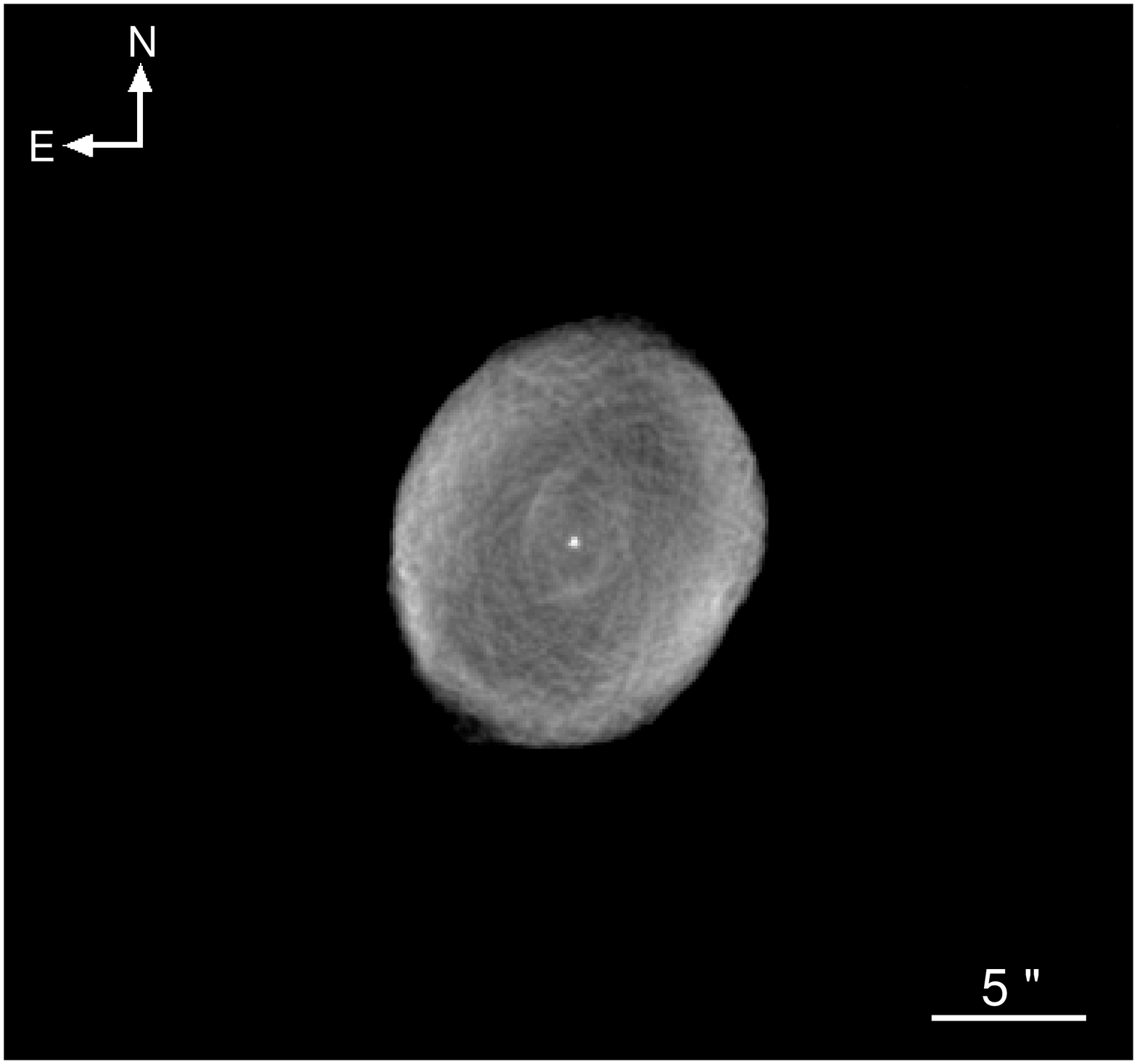}
\vskip .1in
\includegraphics[width=0.9\linewidth]{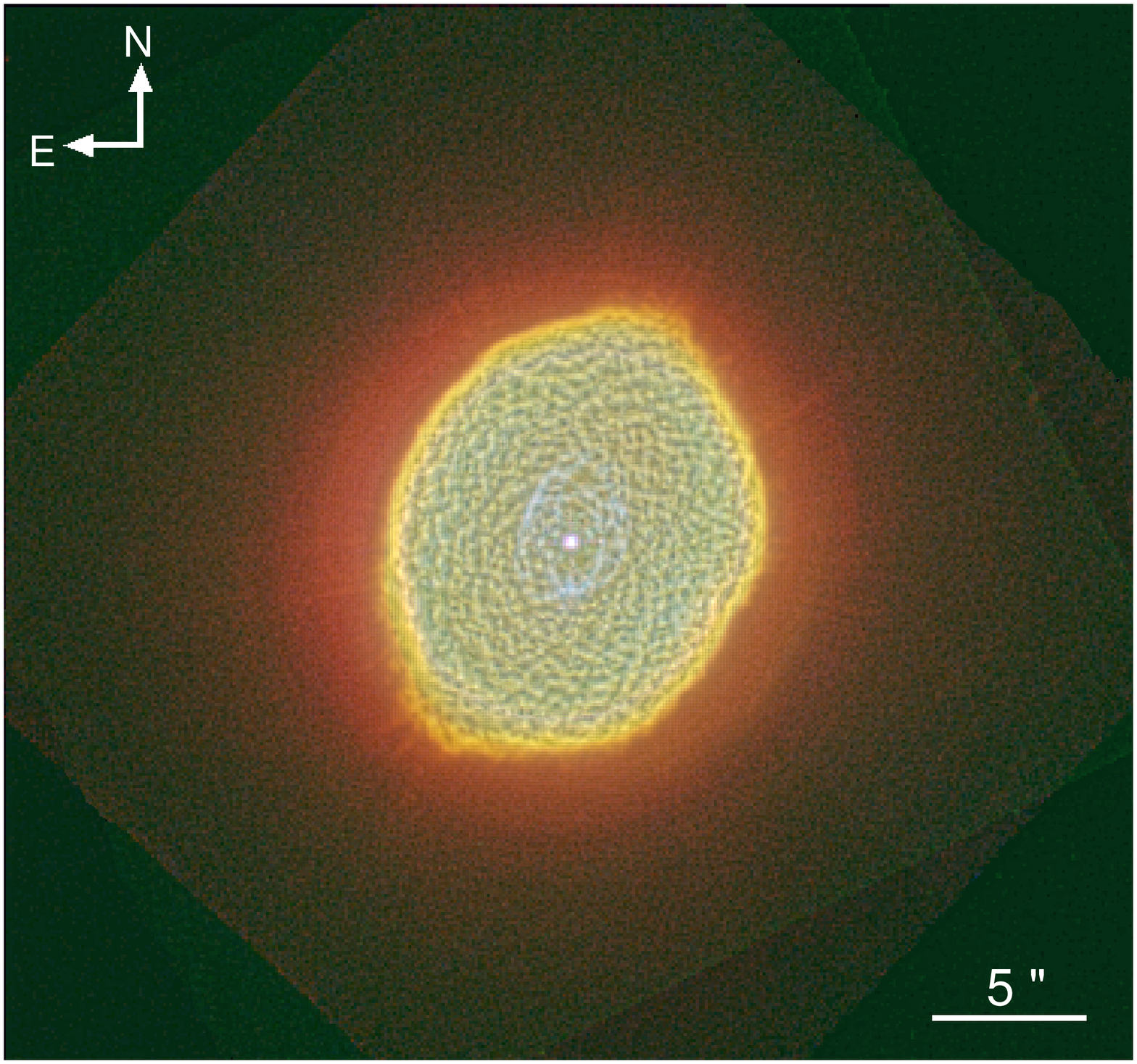}
\vskip .1in  
\caption{TNG $J$ {\it (top)}, \emph{HST} H$\alpha$ {\it (middle)}, and \emph{HST} 
colour composite {\it (bottom)} pictures of IC\,418.  
The \emph{HST} colour composite picture has been processed using unsharp 
masking techniques and includes the [O~{\sc iii}] $\lambda$5007 (blue), 
H$\alpha$ (green), and [N~{\sc ii}] $\lambda$6583 (red) narrow-band 
images.}
\label{HST.img}
\end{figure}

\subsection{Long-slit optical spectroscopy}

Long-slit high dispersion optical spectroscopy of IC\,418 was 
acquired on December 2010 and February 2011 using the Manchester Echelle Spectrometer 
\citep[MES,][]{2003RMxAA..39..185M} at the f/7.5 focus of the OAN-SPM 2.1m 
telescope.  
Since the spectrometer has no cross-dispersion, a $\Delta\lambda$=90\AA\ 
bandwidth filter was used to isolate the $87^{\rm th}$ order of the spectrum 
covering the H$\alpha$ and [N~{\sc ii}] $\lambda\lambda$6548,6583\AA\ 
lines.
The $2048\times2048$ Thomson CCD with a 
pixel size of $15\mu$m was used, resulting a plate scale of 
$0\farcs352\,{\rm pixel}^{-1}$ and a dispersion of 0.06\,{\AA}\,pixel$^{-1}$.

Two slits at PA=0{\degr} and PA=275{\degr}, crossing the central star, were taken with an exposure time of 120\,s.
The data were bias-substracted and wavelength calibrated using a ThAr lamp, with a precision in velocity of $\pm1$\,km\,s$^{-1}$. The seeing during the observations, as determined from the 
FWHM of stars in the FoV, was 1\farcs0.

\begin{figure*}
\begin{center}
\includegraphics[width=1.6\columnwidth]{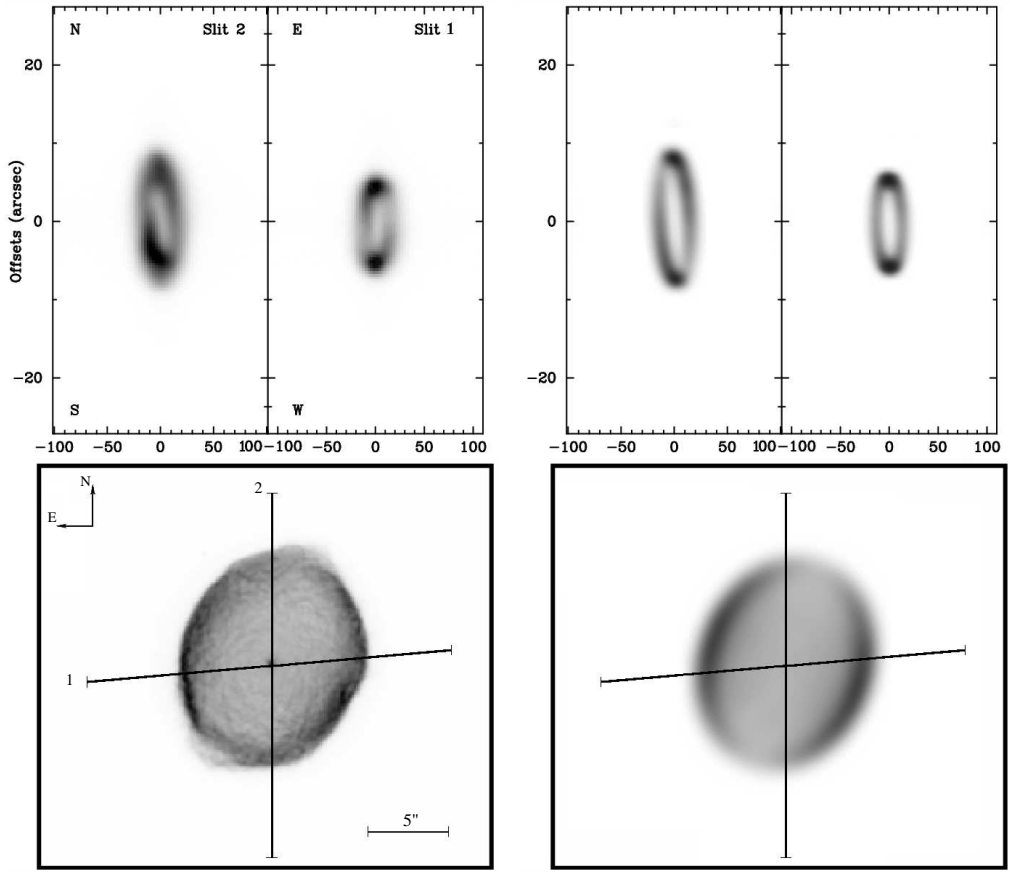}
\caption{ {\it Left} ({\it top}) Position-velocity (PV) maps of the inner shell of IC\,418 and ({\it bottom}) {\it HST\/} WFPC2 [N~{\sc ii}] image overplotted by the positions of the long-slits used for the Echelle spectroscopy. The slits orientation are labeled on the panels. {\it Right} ({\it top}) Synthetic SHAPE (PV) maps of the IC\,418 core and ({\it bottom}) SHAPE model of IC\,418.}
\label{Echelle.img}
\end{center}
\end{figure*}

Low dispersion, long-slit optical spectra of IC\,418 was obtained on two observational runs.
The first on 2011 April 2, using the Boller \& Chivens spectrometer mounted at the 
prime focus of the 2.1 m telescope at the Observatorio Astron\'omico 
Nacional, San Pedro M\'artir, Baja California, Mexico. 
The Marconi 2048$\times$2048 CCD was used as a detector, in conjunction with 
a 400 l mm$^{-1}$ grating blazed at 5500 \AA.  
The slit had a length of 5\arcmin\ and a width of 200 $\mu$m 
($\equiv$2\arcsec).  
The plate and spectral scales were 0\farcs57~pixel$^{-1}$ and 1.7~\AA~pixel$^{-1}$, 
respectively. The spectral resolution was $\sim$4 \AA, the wavelength uncertainty was $\sim$1 \AA, and the spectral range covered was 4080--7560 \AA. 
One 1800\,s exposure was obtained with the slit oriented along the East-West direction and offset by 23\arcsec\ North from the central star.  
The mean air mass during the observations was $\simeq$1.4.  
The observations were flux calibrated using a 600~s exposure of the star 
Hz\,44 obtained with an airmass of 1.03 on the same night.
The seeing as determined from the FWHM of stars in the FoV, was 
$\sim$2\farcs5.

Additional low-dispersion spectra of IC\,418 were obtained on 2011 October 
4, using the Albireo spectrograph at the 1.5 m telescope of the Observatorio 
de Sierra Nevada (OSN), Granada, Spain.  
The Marconi 2048$\times$2048 CCD was used in conjunction 
with a 400 l mm$^{-1}$ grating blazed at 5500 \AA. 
The slit length was $\sim$6\arcmin\ and its width was set at 50 $\mu$m  
($\equiv$2.5\arcsec).  
The binning 2$\times$2 of the detector implied plate and spectral 
scales of 0\farcs30~pix$^{-1}$ and 1.89~\AA~pix$^{-1}$, respectively. 
The spectral resolution was $\sim$4.7 \AA, the wavelength-calibration 
uncertainty $\sim$1 \AA, and the spectral range covered 
3500--10000 \AA. Two positions with exposures of 900 and 1800 seconds were obtained with the slit oriented along the North-South direction (P.A.=0\degr) and offset 
by 65\arcsec\ and 105\arcsec\ West from the central star, respectively.  
The observations were flux calibrated using spectra of the 
spectrophotometric standard star Hiltner~600 acquired on the 
same night.  
The seeing, as determined from the FWHM of stars in the FoV, 
was $\sim$2\farcs5.

All intermediate-dispersion spectra were bias-subtracted, flat-fielded, and wavelength and flux 
calibrated following standard procedures using {\small XVISTA}\footnote{
{\small XVISTA} is maintained and distributed by Jon Holtzman at the New Mexico 
State University\\ http://astro.nmsu.edu/$\sim$holtz/vista}, {\small IRAF} 
and the nebular analysis software {\small ANNEB} \citep{2011RMxAC..40..193O}
which integrates the {\small NEBULAR} package of {\small IRAF/STSDAS} \citep{1995PASP..107..896S}.

\section{Multiple shells around IC\,418}\label{sec_res}

The new images of IC\,418 reveal a wealth of structures around its 
ionised bright shell. 
These are described into further detail below. 

\begin{figure*}
\centering
\includegraphics[width=0.85\columnwidth]{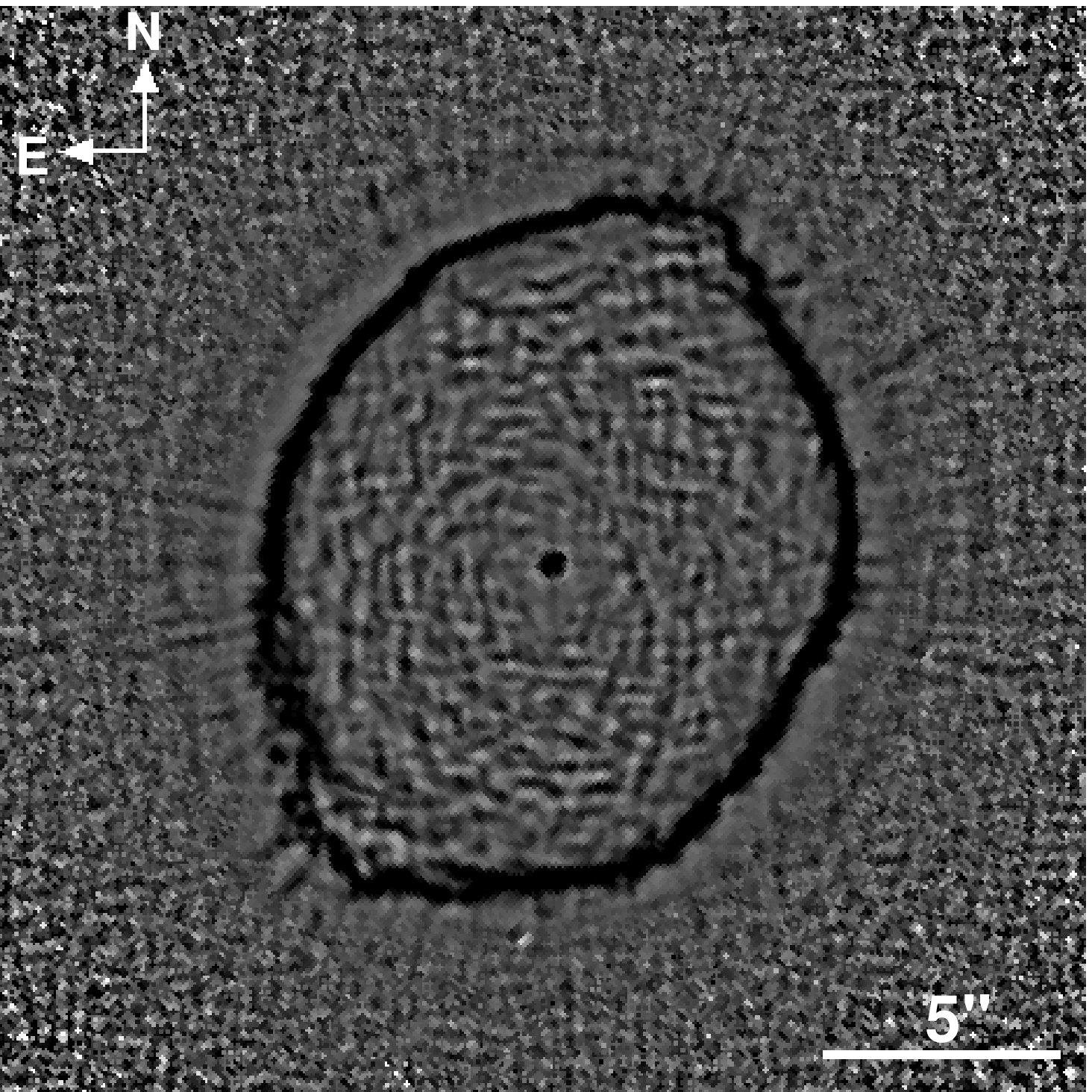}
\hspace*{\columnsep}
\includegraphics[width=0.85\columnwidth]{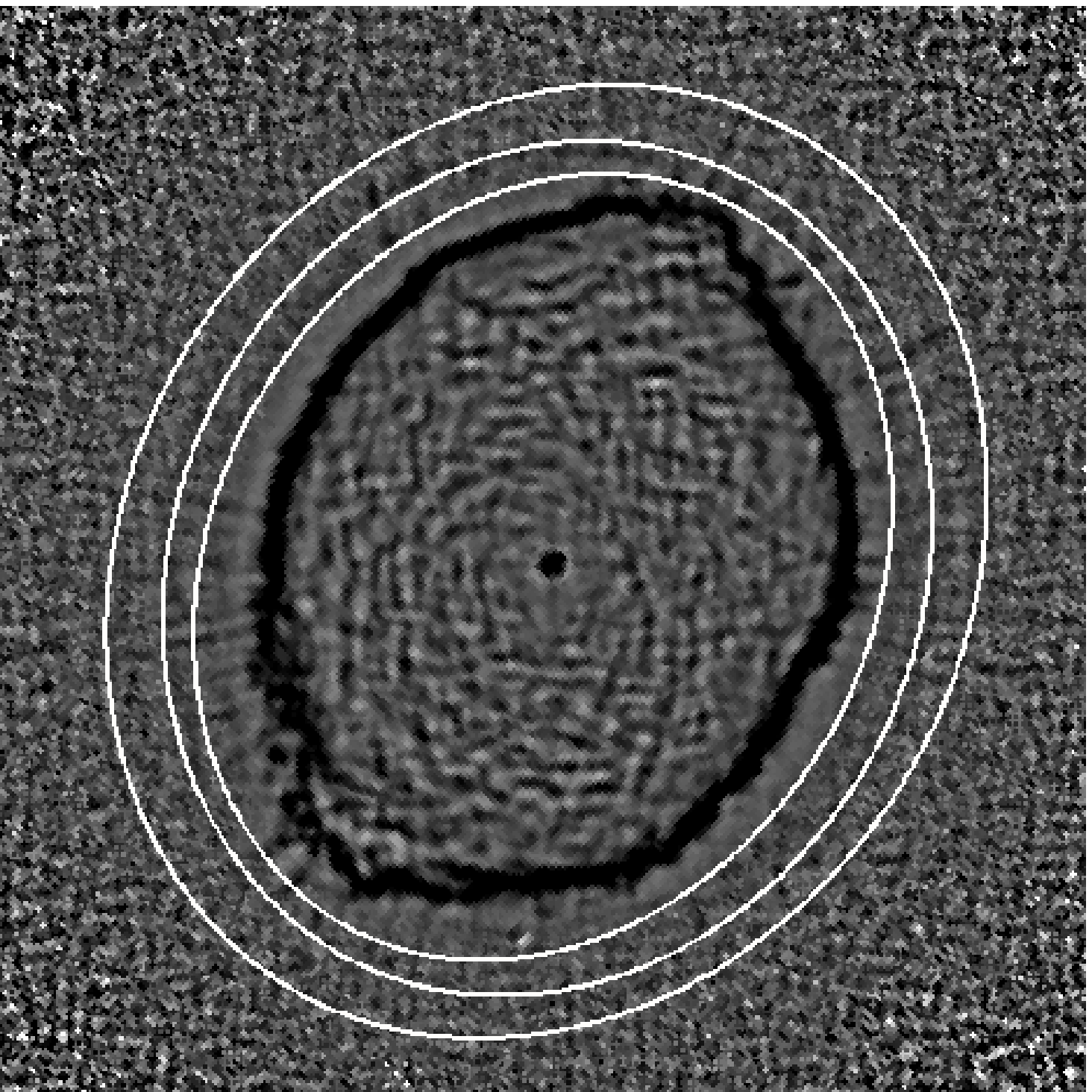}
\caption{\emph{HST} [N~{\sc ii}] image of IC\,418 processed to emphasize the ring 
features around the bright inner shell following the techniques developed 
by \citet{2004AA...417..637C}. This process has also enhanced a series of radial filaments in the image. 
The left panel corresponds to the processed image, where darker shades 
of grey indicate higher levels of emission.  
The location of the rings has been overlaid by white arcs in the 
right panel.}  
\label{rings.img}
\end{figure*}

\subsection{The inner shell}

Detailed morphological descriptions of the inner shell of IC\,418 abound 
in the literature \citep[see, for instance, the recent work by][and 
references therein]{2011AJ....141..134S}.  
We will mention here the distinctive pattern of interwound features 
(Figure~\ref{HST.img}-{\it bottom}), and the 3\farcs5$\times$5\farcs5 
shell interior to the 11\arcsec$\times$14\arcsec outer shell.  
This inner shell is particularly bright in [O~{\sc iii}], fainter in 
H$\alpha$, and undetected in [N~{\sc ii}], implying a higher relative 
excitation, a characteristic which is common of other PNe with these 
inner bubbles \citep{2011AJ....141..134S}.  
Finally, we note the morphological similarities between the H$\alpha$ 
and near-IR $J$ images (Figure~\ref{HST.img}-{\it top,middle}) which 
suggests that the $J$ band image is dominated by the H~{\sc i} Pa$\beta$ 
$\lambda$1.2817 $\mu$m line \citep{1999ApJS..124..195H}.

It is noticeable that, in spite of the high brightness of IC\,418 and the 
substantial number of studies in the literature about its morphology and 
physical conditions, a spatio-kinematical study of its three-dimensional 
structure is lacking.  
We next describe a simple morpho-kinematical model that fits the 
observations described in $\S$2.

Figure~\ref{Echelle.img} ({\it bottom-left}) presents the \emph{HST}
image of the inner shell overplotted with the two slits going across the central
star. 
The corresponding [N~{\sc ii}] PV maps of the inner shell are presented 
in Figure~\ref{Echelle.img} ({\it top-left}).  
We have used the computational tool SHAPE \citep{2010arXiv1003.2012S} to 
construct a simple 3D model of IC\,418 by fitting simultaneously 
the [N~{\sc ii}] \emph{HST} image (morphology) and PV maps 
(kinematics). 
The best model consists of an ellipsoid with semi-major axis of 6\farcs6 
and semi-minor axis of 5\farcs3 oriented along PA$=341{\degr}$ and with 
an inclination angle of $i=65{\degr}$ with respect to the line of sight. Assuming homologous expansion, we derive a polar velocity $V_{\rm pol}=15.7$~km\,s$^{-1}$ and an equatorial velocity $V_{\rm eq}=12.6$~km\,s$^{-1}$, notably larger than the 
expansion velocity of 5~km\,s$^{-1}$ recently reported by 
\citet{2012IAUS283}, but consistent with the value of 
12~km\,s$^{-1}$ provided by \citet{GAS96}.  
The kinematical age of the model, for the distance of 1.3 kpc, is 
$\tau_{\rm kin}=2600 \pm800\,{\rm yr}$.  

We note that the model and kinematical age discussed above only applies to 
the outer skirts of the shell probed by the emission in the low-ionization 
[N~{\sc ii}] line.  
Higher excitation material exhibits lower expansion velocities, as 
suggested by the unresolved H$\alpha$ emission line in our echelle 
observations and more reliably by the unresolved profile of the 
otherwise narrow [O~{\sc iii}] emission line presented by \citet{GAS96}.  
Since higher excitation material is closer to the central star, 
the difference in line profiles reveals a velocity gradient 
throughout the nebula.

\begin{figure}
\centering
\includegraphics[width=0.99\linewidth]{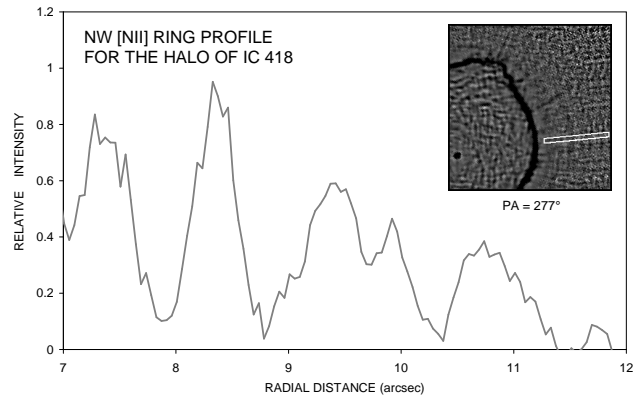}
\caption{Profile through the northwestern region (PA = 277\degr\,) halo ring structures, where the width of the slice is $\sim$7.3 pixels ($\equiv$0.33\arcsec). Underlying halo emission has been removed using a fifth-order least-squares polynomial fit.}  
\label{ringprofile.img}
\end{figure}

\begin{table*}
\begin{center}
\begin{minipage}{142mm}   
\caption{Line ratios comparison for the OSN and OAN observations of IC\,418.}
   \begin{tabular}{@{}lccccccccccc@{}}
   \hline
    Line ratio & \multicolumn{2}{c}{OSN} & OAN & P04 & H00 & H94 & JPP90 & GM85 & A83 & TP80 & K76 \\
\cline{2-3}
\cline{4-4}
\cline{8-8}
  & W65$\arcsec$ & W105$\arcsec$ & N23$\arcsec$ & & & & 15$\arcsec$ & & & &\\
  \hline
[O~{\sc iii}]$\lambda$5007/H$\alpha$&0.69&0.52&0.44&0.36&0.43&0.24&0.32&0.53&0.53&0.46&0.49~\\
\~[N~{\sc ii}]$\lambda$6584/H$\alpha$&0.41&0.67&0.63&0.64&0.61&0.59&0.66&0.53&0.57&0.59&0.61~\\
 \hline
 \end{tabular}
\textbf{Notes:}
OSN (Observatorio de Sierra Nevada) this paper; 
OAN (Observatorio Astron\'omico Nacional) this paper; 
P04: Pottasch, S.R., et al., (2004); 
H00: Henry, R.B.C., et al., (2000); 
H94: Hyung, S., et al., (1994);
JPP90: Phillips, J.P., et al. (1990);
GM85: Guti\'errez-Moreno, A., et al, (1985); 
A83: Aller, L.H. \& Czyzak S.J., (1983);
TP80: Torres-Peimbert, et al., (1980);
K76: Kaler, J.B., (1976).
\end{minipage}
\end{center}
\end{table*}

\subsection{Radial filaments and rings}

The \emph{HST} [N~{\sc ii}] image and, to some extent, the H$\alpha$ image, 
disclose a system of radial filaments or rays that emanate from the bright 
shell of IC\,418, as illustrated by the \emph{HST} composite-colour picture 
shown in Figure~\ref{HST.img}-{\it bottom}. 
These filaments are mostly radial, pointing towards the location of the 
central star, and their surface brightness is small, $\leq$2\% the surface 
brightness peak of the main nebula. 
Their size is variable, but in a few cases we measure extensions 
up to 2\farcs7.  
We also note the presence of little blisters or ``bubble-like'' features 
onto the exterior walls of the bright nebular shell.

A close inspection of the regions just outside the bright ionised shell of 
IC\,418 reveals [N~{\sc ii}] and H$\alpha$ emission distributed along several elliptical arcs. 
These are best seen in Figure~\ref{rings.img}, where we can identify 
up to three concentric elliptical arcs that enclose the inner shell. 
Their surface brightness is significantly low, $\sim$2.5 \%, $\sim$1.4 \%, 
and $\sim$0.7 \% the brightness peak of the inner shell, and their separation 
can be estimated to be $\sim$1\arcsec. Although these are among the observations with the highest-resolution so far obtained for structures of this kind \citep[see also][]{2001AJ....121..354B,2004AA...417..637C}, they show the rings to be featureless and relatively broad.   
A profile through the northwestern region of these structures is shown in Figure~\ref{ringprofile.img}. The surface-brightness profile is dominated by emission scattered from the bright inner shell that we have subtracted using a fifth-order least-squares polynomial. Figure~\ref{ringprofile.img} suggests that the separation between rings is no uniform, with an averaged distance between peaks $\sim$1\arcsec, similar to our previous estimate. This profile even suggests a larger number of rings, but the quality of the observations precludes us to make a more definite statement.

\subsection{Haloes}

\subsubsection{Inner halo}

\begin{figure}
\begin{center}
\includegraphics[width=0.9\linewidth]{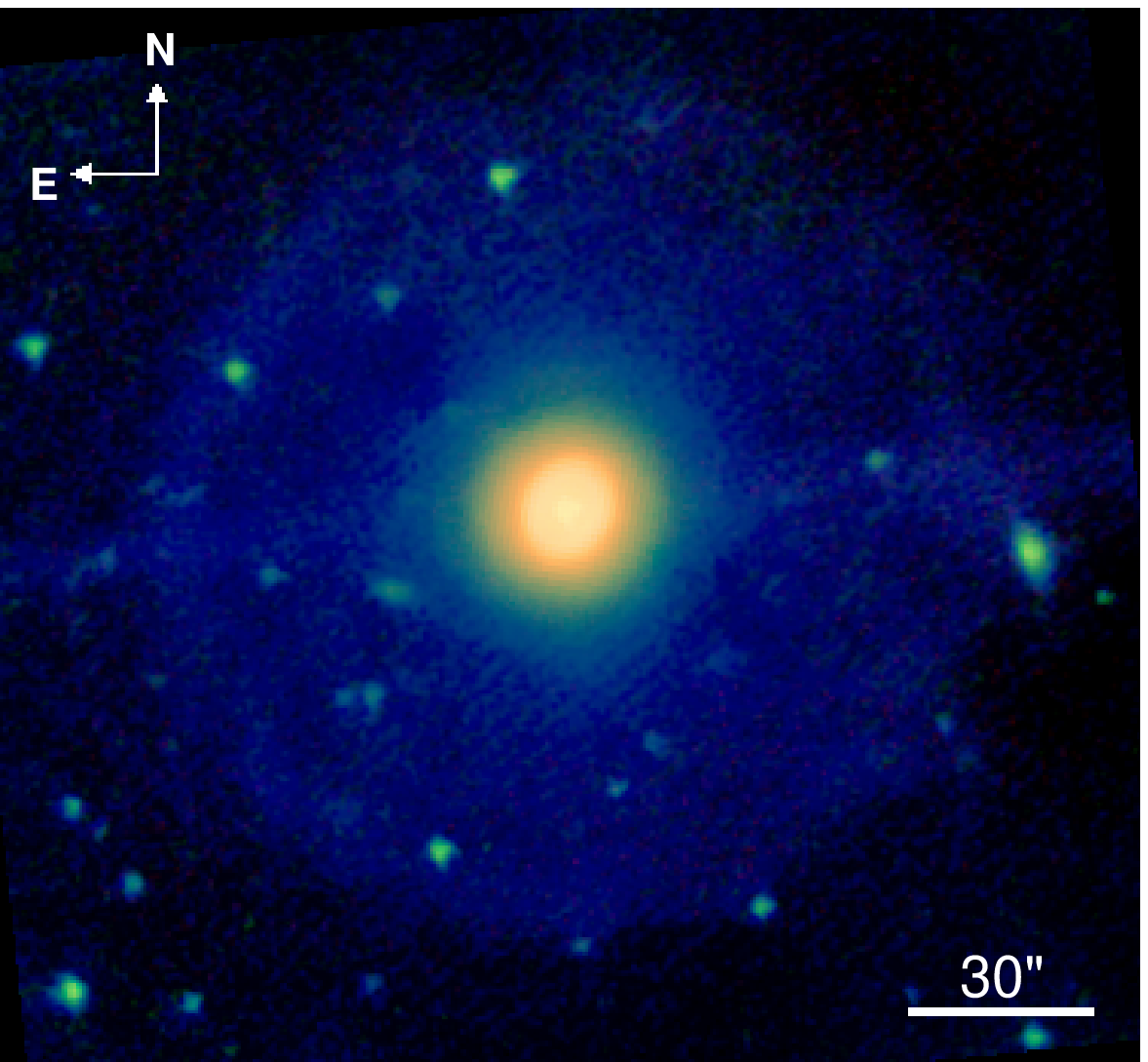}
\vskip .1in
\includegraphics[width=0.9\linewidth]{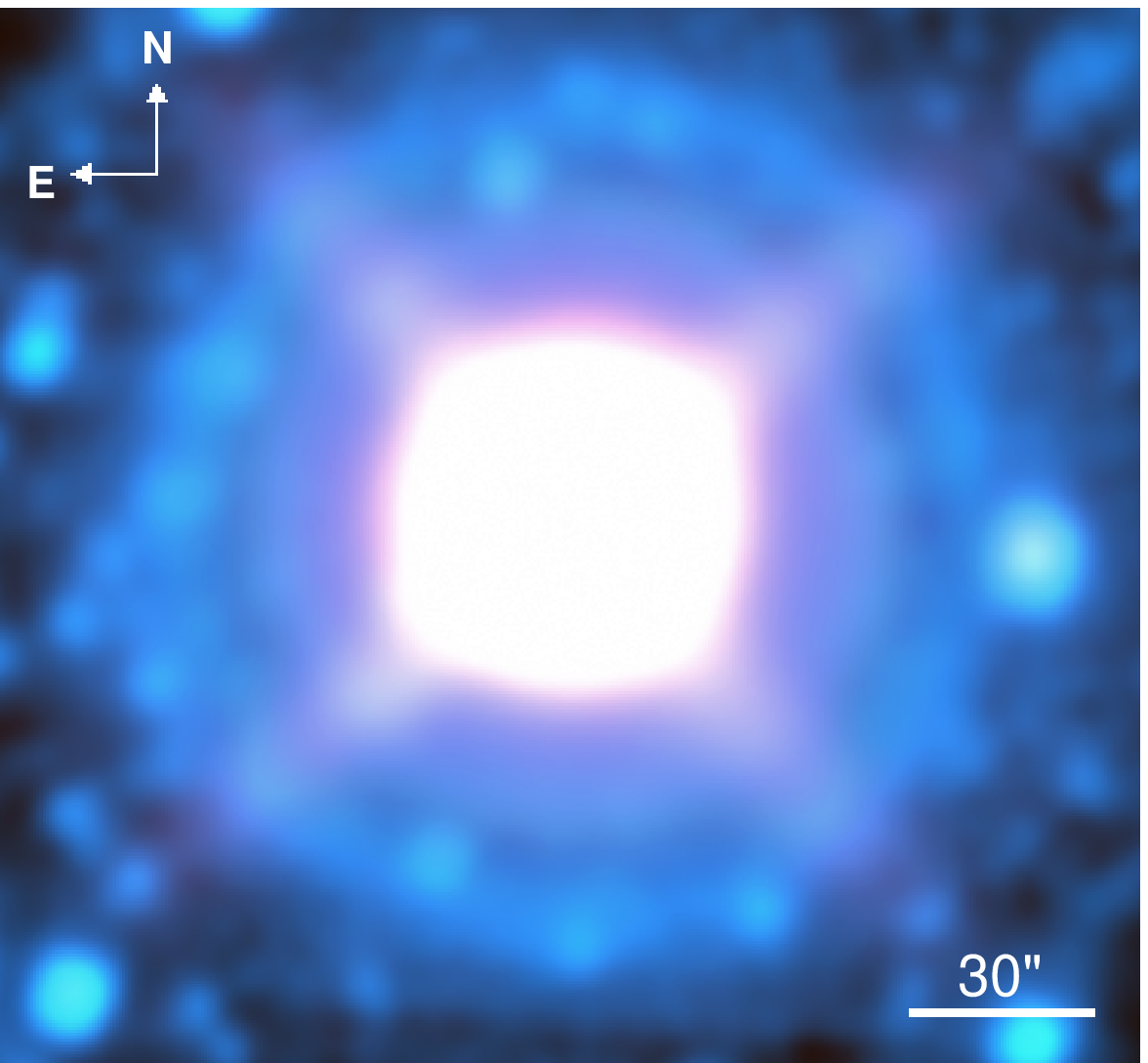}
\vskip .1in
\includegraphics[width=0.9\linewidth]{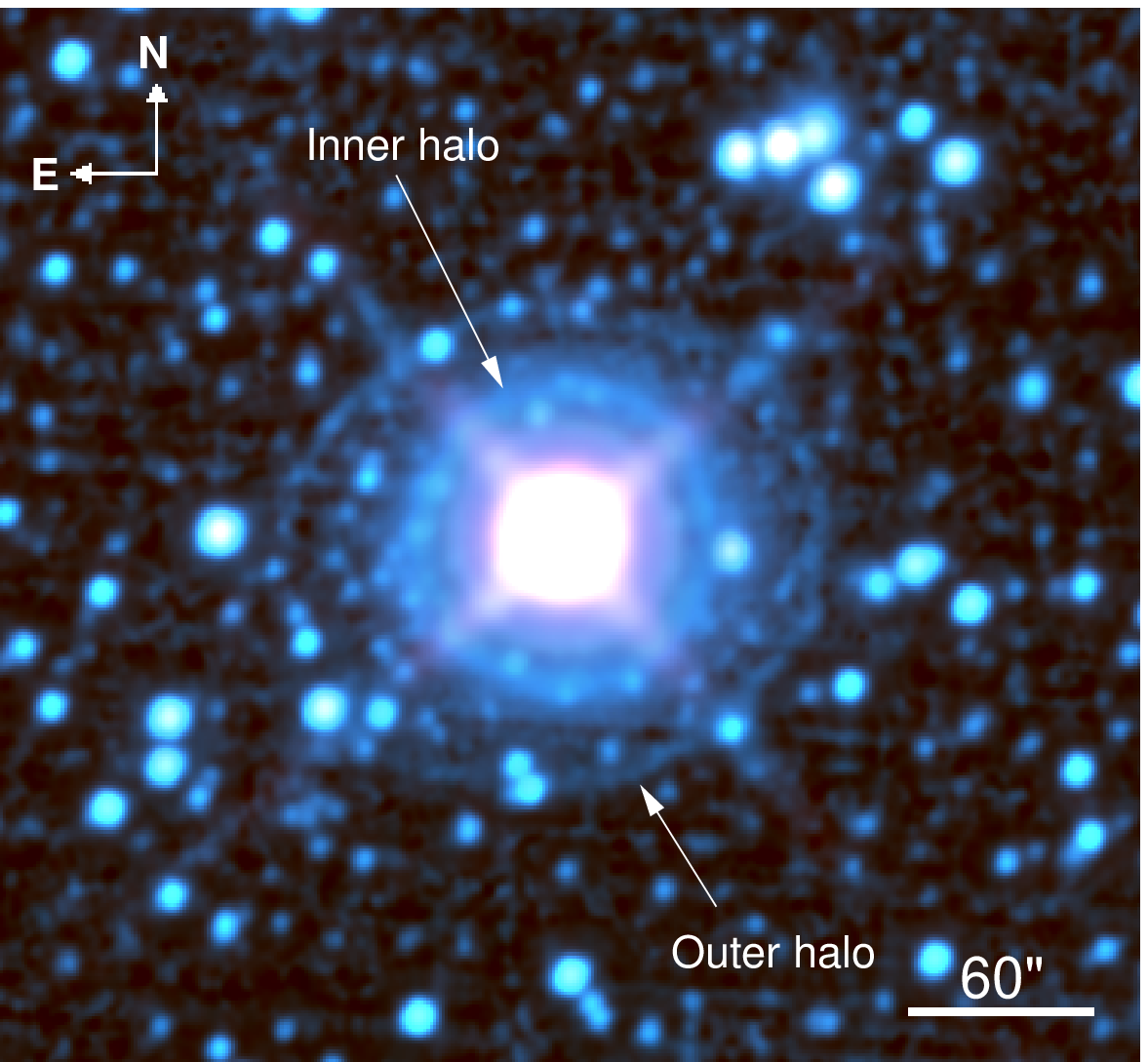}
\caption{({\it top}) OAN CAMILA $JHK$ and ({\it middle-bottom}) 
\emph{WISE} (W1 and W2) composite colour pictures of IC\,418.
In the $JHK$ picture, $J$ is blue, $H$ is green, and $K'$ is red, 
while in the \emph{WISE} picture, W1 is blue-green and W2 is red. 
The \emph{WISE} images have been processed using unsharp masking 
techniques, leading to the apparent image ``sharpening'' 
\citep{Levi1974}.  
}  
\label{wise.img}
\end{center}
\end{figure}

After the publication of the analysis of \emph{J$-$H} and \emph{H$-$K$_S$} 
maps of a sample of PNe including IC\,418 \citep{2005MNRAS.364..849P}, 
a close inspection of the \emph{2MASS} \emph{J}-band image hinted the 
presence of a round faint halo with radius $\simeq$75\arcsec. 
Since the detection of faint structures around bright nebulae is hindered by 
instrumental artifacts that produce ``ghost haloes'' \citep{2003MNRAS.340..417C}, 
the true nature of this structure remained unclear. 
The new $JHK$ band images obtained with CAMILA 
(Figure~\ref{wise.img}-{\it top}) find a halo 
with the same spatial extent as the one hinted 
in \emph{2MASS} images. 
The additional \emph{WISE} images unambiguously confirm the presence and 
spatial extent of this halo that will be referred as the inner halo of 
IC\,418.

This inner halo is basically detected in the $J$ near-IR band and the W1 
3.4 $\mu$m \emph{WISE} band (Figure~\ref{wise.img}-{\it middle}), with much fainter emission in the $H$ and $K$ 
near-IR and the W2 \emph{WISE} bands. 
The nature of this emission is clarified by the near-IR spectrum of 
the halo of IC\,418 described by \citet{1999ApJS..124..195H} that shows the H~{\sc i} Pa$\beta$ $\lambda$1.2817 $\mu$m line to be the main contributor 
to the emission in the $J$ band.  
Therefore, the inner halo of IC\,418 is ionised.  
There are no mid-IR spectra for the halo, but the \emph{ISO} spectrum 
of the central region of IC\,418 presented by \citet{2004AA...423..593P} 
suggests that the \emph{WISE} W1 band may be dominated by the H~{\sc i} 
5-4 $\lambda$4.053 $\mu$m line. 
The lack of continuum emission in the \emph{ISO} spectrum further strengthens 
our presumption that the emission from this halo comes from ionised material.

\subsubsection{Outer halo}

The \emph{WISE} images of IC\,418 reveal an unexpected additional 
elliptical shell with its major axis along PA$\sim$300\degr\ and 
a size of 220\arcsec$\times$250\arcsec (Figure~\ref{wise.img}-{\it bottom}).  
This shell is $\sim$5 times fainter than the inner halo in the 
\emph{WISE} W1 band, it has a noticeable limb-brightened morphology, 
and its Western tip is brighter than the Eastern tip.  
Given its limb-brightened morphology, we will refer to it as the outer 
halo of IC\,418 \citep{1987ApJS...64..529C}.   

The elliptical morphology and asymmetrical brightness distribution of this 
outer halo may be indicative of the motion of IC\,418 relative to the 
interstellar medium \citep{1996ApJS..107..255T, 2003ApJ...585L..49V,2007MNRAS.382.1233W}. The off-centered 
position of the main nebula, shifted by 6\arcsec\ along the direction 
pointing towards the apex of the outer halo at PA$\sim$300\degr, is 
consistent with this hypothesis.

\subsubsection{{\bf Spectral properties of the haloes}}

High- and intermediate-dispersion long-slit spectroscopic observations of the haloes of IC\,418 have been obtained in an effort to determine their kinematical and spectral properties. Unfortunately, no kinematical information is available because the low surface brightness of the haloes has not allowed us to detect their emission in the MES high-dispersion spectroscopic observations, but the OAN and OSN intermediate-dispersion spectroscopic observations have detected the emission from these haloes. The values of the [O~{\sc iii}] $\lambda$5007/H$\alpha$ and [N~{\sc ii}] $\lambda$6584/H$\alpha$ line ratios derived from these observations are listed in Table~2, together with the values reported by different authors for the central region of IC\,418.

A statistical analysis of the different values of the line ratios reported by different authors for the central region of IC\,418 yields averaged [O~{\sc iii}] $\lambda$5007/H$\alpha$ and [N~{\sc ii}] $\lambda$6584/H$\alpha$ line ratios of 0.42$\pm$0.10 and 0.60$\pm$0.04, respectively. These lines ratios are analogous to those derived from the OAN spectrum offset 23\arcsec ~North, 
marginally consistent with those of the OSN spectrum offset 105\arcsec ~West, and notably different to those of the OSN spectrum offset 65\arcsec ~West. We note here that the OAN spectrum is too close to the bright central nebula and, as a result, it is considerably contaminated by dispersed light. The OSN spectrum offset 105\arcsec ~West is very faint and the line ratios affected by considerable uncertainties. On the other hand, the OSN spectrum offset 65\arcsec ~West shows definite different line ratios than the central regions of IC\,418 and the spatial distribution of the emission is consistent with the limb-brightened morphology of this halo, proving that the inner halo is indeed ionized and that its excitation is higher than this of the inner regions of IC\,418.

\section{Discussion}\label{sec_dis}

IC\,418 is a bright and young PN which is optically thick to H-ionising radiation.  
Such condition may imply the existence of a region of neutral or molecular 
material outside the main nebula, although conclusive evidence for the 
occurrence and spatial extent of this region has been lacking.  
Using narrow-band optical and broad-band near-IR and mid-IR images, 
and optical long-slit intermediate-dispersion and echelle spectra, 
we have searched for emission outside the bright inner 
shell of IC\,418.  
The analysis of these observations has revealed different structures 
outside the main nebular shell of IC\,418: radial rays, rings, blisters and a 
double system of haloes.

These features are common in many other 
extended PNe.  
For instance, the blisters, ``bubble-like'' features onto the surface of the inner shell can be seen in the Ring Nebula, 
NGC\,6720 \citep{2002AJ....123.3329O} and they reveal that the outer walls of the ionised shells 
of PNe may be subject of turbulence and instabilities.  
Meanwhile, the radial filaments \citep[rays in the nomenclature of] []{2004AJ....127.2262B,2011AJ....141..134S}, have been described in many PNe such as NGC\,6853 \citep{2006ApJ...652..426H}, NGC\,6543 \citep{2004AJ....127.2262B}, or in the mid-IR observations of NGC\,40 
\citep{2011MNRAS.411.1245R}.  
The origin of these features has been associated to ``ionisation shadows'' 
produced by dense knots and the subsequent ionisation by leaking UV photons \citep{2004AJ....127.2262B}. 

The rings around the bright central regions of PNe have been reported for a growing 
number of PNe \citep[e.g., ][]{2004AA...417..637C,2005ApJ...635L..49K,2011AJ....141..134S,
2011MNRAS.411.1245R}. 
Their origin and formation are uncertain \citep[see, for instance, 
][]{2000ApJ...540..436S,2001ApJ...560..928G}, but they represent radial enhancements in particle density that suggests mass-loss gasping in 
the late phases of the AGB evolution. 
At the distance of 1.3 kpc towards IC\,418 \citep{2009AJ....138...46G}, 
the time-lapse between rings would be $\sim630/(v/\mathrm{10 km s}^{-1})$ yr, 
where a typical expansion velocity of 10 km~s$^{-1}$ is assumed for the AGB wind.
\\

The double-halo morphology of IC\,418 in the IR is very similar to 
those found in Cn\,1-5, IC\,2165, NGC\,2022, and NGC\,6826 in the optical 
\citep{2003MNRAS.340..417C}, possibly implying similar formation mechanisms. 
\citet{2003MNRAS.340..417C} argue that the inner of the two haloes of 
these PNe is a recombination halo \citep{1999MNRAS.302L..17R}, whereas the outer halo is an AGB halo.
This is not probably the case for IC\,418, as the ionising flux of 
its central star has not reached its maximum value yet.

The time-lapse between shells of multiple shell PNe can be compared 
to the time between thermal pulses at the end of the AGB phase 
\citep{VW93}, although hydrodynamical and radiative processes can 
alter appreciably the kinematics and evolution of the haloes and 
limit this comparison \citep{VGM02,VMG02,SS03}. At the distance of 1.3 kpc and adopting a typical expansion velocity of 
10 km~s$^{-1}$, the kinematical ages of the outer and inner halo are 
$\sim69,000/(v/\mathrm{10 km s}^{-1})$ and $\sim47,000/(v/\mathrm{10 km s}^{-1})$ yr, 
respectively, whereas the kinematical age of the inner shell was 
estimated to be $\tau_{\rm kin}=2600\,\mathrm{yr}$.  
The inter-shell time lapses are, thus, $\sim20,000/(v/\mathrm{10 km s}^{-1})$ yr 
and $\sim45,000/(v/\mathrm{10 km s}^{-1})$ yr.  
Accounting for the lack of information on the expansion of the haloes 
of IC\,418 and for the very likely dynamical effects that have modified 
their kinematical age, the mass-loss history of IC\,418 can be roughly described to consist of three major episodes of mass-loss separated by 10,000--50,000 yr, with the final one occurring $\sim$3,000 yr ago after having been preceded by several 
mass-loss gasps. \\

In the detection of the faint haloes around IC\,418, the near- and mid-IR 
images have provided especially helpful.  
Indeed, the generalized use of near- and mid-IR observations of PNe 
have detected a significant number of faint outer structures around 
the main nebular shells of many PNe, e.g., 
MeWe\,1-3 and NGC\,6852 \citep{2009AJ....138..691C}, 
NGC\,40 \citep{2011MNRAS.411.1245R}, 
NGC\,3242 and NGC\,7354 \citep{2009MNRAS.399.1126P}, 
NGC\,1514 \citep{2010AJ....140.1882R}.

\section{Summary}\label{sec_con}

We present observations of IC\,418 that for the first time probe the 
presence of different structures around the bright main nebular shell 
including radial filaments or rays, a system of three concentric rings, 
and two detached haloes.  
The outer halo has an elliptical shape and the main nebula is off-centered 
with respect to this halo. 
 
The inter-shell time laps between the two haloes and the main nebula are derived to be 10,000--50,000 yr, and the time-lapse between the three 
concentric rings is $\sim$630 yr for an assumed expansion velocity of 10 
km~s$^{-1}$. These concentric rings can be associated to several episodes of mass-loss gasps that preceded the final mass-loss event that formed the main nebular shell.

Finally, we note that the progression of the ionisation front through 
the nebula is not homogeneous, with the development of instabilities 
at the outer regions of the ionised shell and the formation of radial 
structures caused by ``shadows'' to the leaking UV photons.

\section*{Acknowledgments}

GRL acknowledges support from CONACyT and PROMEP (Mexico).
MAG and GRL are partially funded by grant AYA2008-01934 of the 
Spanish Ministerio de Ciencia e Innovaci\'on (MICINN) which includes FEDER funds.
RV, LO, MAG, and GRL thank support by grant IN109509 (PAPIIT-DGAPA-UNAM). 
MAG also acknowledges support of the grant AYA 2011-29754-C03-02. Also we want to thank to the OAN-SPM staff and the CATT for time allocation and an anonymous referee who made very useful comments for the improvement of the paper.

Based on observations made with the 2.1m telescope of the Observatorio Astron\'omico Nacional at the Sierra de San Pedro M\'artir (OAN-SPM), which is a national facility operated by the Instituto de Astronom\'{\i}a of the Universidad Nacional Aut\'onoma de M\' exico, the Italian Telescopio Nazionale Galileo (TNG) operated on the island of La Palma by the Fundaci\'on Galileo Galilei of the INAF (Istituto Nazionale di Astrofisica) at the Spanish Observatorio del Roque de los Muchachos of the Instituto de Astrof\'{\i}sica de Canarias, and the 1.5m telescope at the Observatorio de Sierra Nevada (OSN), Granada, Spain, which is operated by the Instituto de Astrof\'{\i}sica de Andaluc\'{\i}a.

This research has made use of the NASA/IPAC Infrared Science Archive, which is operated by the Jet Propulsion Laboratory, California Institute of Technology, under contract with the National Aeronautics and Space Administration. We have also used archival observations made with the NASA/ESA Hubble Space Telescope, and obtained from the Hubble Legacy Archive, which is a collaboration between the Space Telescope Science Institute (STScI/NASA), the Space Telescope European Coordinating Facility (ST-ECF/ESA) and the Canadian Astronomy Data Centre (CADC/NRC/CSA). 

We would like to dedicate this paper in memory of our colleague and friend Dr.\ Yolanda G\'omez Castellanos who recently passed away. We would also like to 
remember Prof.\ John Peter Phillips whose inspiration provided the motivation for this research although he unfortunately could not see it complete.

%\bsp

\label{lastpage}

\end{document}